\documentclass[3p,times]{elsarticle}
\usepackage{ecrc}
\volume{00}
\firstpage{1}
\journalname{Nuclear Physics A}

\runauth{R. V\'ertesi (PHENIX collaboration)}

\jid{nupha}

\usepackage{amssymb}
\usepackage{amsthm}
\usepackage{lineno}

\usepackage{graphicx}
\usepackage{amssymb}
\usepackage{epstopdf}
\usepackage{geometry}  
\usepackage{xspace}

\usepackage[figuresright]{rotating}

\newcommand{\snn}{\ensuremath{\sqrt{s_\mathrm{NN}}}\xspace}
\newcommand{\Raa}{\ensuremath{R_\mathrm{AA}}\xspace}

\newcommand{\Ncoll}{\ensuremath{N_\mathrm{coll}}\xspace}
\newcommand{\Npart}{\ensuremath{N_\mathrm{part}}\xspace}
\newcommand{\piz}{\ensuremath{\pi^0}\xspace}
\newcommand{\pt}{\ensuremath{p_\mathrm{T}}\xspace}
\newcommand{\pT}{\pt}
\newcommand{\Sloss}{\ensuremath{S_\mathrm{loss}}\xspace}
\newcommand{\xt}{\ensuremath{x_\mathrm{T}}\xspace}
\newcommand{\Dphi}{\ensuremath{\Delta\phi}\xspace}
\newcommand{\GeV}{\ensuremath{\mathrm{GeV}}\xspace}

\begin{document}

\begin{frontmatter}


\title{Neutral pion production in Au+Au collisions at RHIC}
\author{R\'obert V\'ertesi for the PHENIX Collaboration}
\ead{vertesi.robert@wigner.mta.hu}
\address{MTA Wigner RCP, RMI, P.O.box 49, 1525 Budapest, Hungary}

\begin{abstract}

New results from the 2010 RHIC low energy program show a substantial suppression of neutral pions in central Au+Au collisions at both \snn=39 and 62.4 GeV c.m.s. energies. At high \pT the 62.4 GeV and 200 GeV data follow the same suppression pattern. On the other hand, otherwise successful pQCD predictions do not describe the 39~GeV data. These observations indicate that initial state effects may play a dominant role at smaller c.m.s. energies and at lower \pT. 
The  azimuthal dependence of the nuclear modification factor \Raa is strongly correlated with the (approximately elliptical) geometry of the overlap region. The dependence of \Raa on the reaction plane, determined up to \pt{}=20 GeV/$c$ from 2007 high-luminosity \snn=200 GeV Au+Au data provides great selectivity among theories, and favours the ASW scenario with AdS/CFT correspondence over the pQCD-based models.

\end{abstract}


\end{frontmatter}


\section{Introduction}

Neutral pion production in \snn=130 and 200 GeV central Au+Au collisions at the Relativistic Heavy Ion Collider has been found to be strongly suppressed in comparison to the expectations from properly scaled p+p collisions~\cite{Adcox:2001jp,Adler:2003qi}, while data from d+Au collisions showed no suppression or enhancement~\cite{Adler:2003ii}, indicating that hadron suppression is a final state effect. This observation was one of the first convincing signatures of a strongly interacting partonic medium created in high energy heavy ion collisions. The nuclear modification factor \Raa quantifies the departure from the above scaling. It is generally defined as a function of transverse momentum and pseudorapidity,
\begin{equation}
\Raa(\pt,\eta) = 
	\frac{ ( 1/N^{evt}_\mathrm{AA} ) \mathrm{d}^2 N^{\piz}_\mathrm{AA} / \mathrm{d}\pT \mathrm{d}\eta }
	{\left< T_\mathrm{AB} \right> \times \mathrm{d}^2 \sigma^{\piz}_\mathrm{pp} / \mathrm{d}\pT \mathrm{d}\eta }	
\end{equation}
where $\sigma^{\piz}_\mathrm{pp}$ is the production cross section of \piz in p+p collisions, 
$\left< T_\mathrm{AB} \right> = \Ncoll / \sigma^{inel}_\mathrm{pp} $ is the nuclear overlap
function averaged over the relevant range of impact parameters, and \Ncoll is the number of binary collisions computed with $\sigma^{inel}_\mathrm{pp}$. 
The \Raa, however, is also subject to initial effects such as multiple soft scattering (Cronin-effect). Measurements of pion production in Cu+Cu collisions at $\snn=22.4, 62.4$ and 200 GeV center-of-mass (c.m.s.) energies have shown that such effects indeed play an important role: while \snn=200 and 62.4 GeV central data in this smaller system still show a suppression, pion production in 22.4 GeV Cu+Cu collisions is moderately enhanced at $\pT>2\ \GeV/c$, virtually independently of centrality~\cite{Adare:2008ad}. Production of \piz{}s has been measured in \snn=62.4 GeV and 39 GeV Au+Au collisions in 2010, as part of the RHIC low energy program~\cite{Adare:2012uk}. These data, which reach up to $\pT=10\ \GeV/c$, allow for the study of the onset and evolution of the suppression, and restricts theoretical models. For a better understanding on the interplay of the different effects of nuclear modification, other observables, such as the fractional momentum shift \Sloss, are also measured.
Azimuthal asymmetries with respect to the reaction plain carry additional information on the collision geometry. Recent PHENIX measurements at $\snn=200\ \GeV$   of \piz azimuth-dependent nuclear modification factor $\Raa(\pT,\Dphi)$ up to $\pT=20\ \GeV/c$~\cite{Adare:2012wg} give additional constraints on energy loss mechanisms.

\section{Experiment and Analysis}

A detailed description of the PHENIX detector is given in~\cite{Adcox:2003en}. 
In the analysis of RHIC year 2007 run $\snn=200\ \GeV$ Au+Au data, $3.8 \times 10^9$ minimum bias events were used. 
Year 2010 run $\snn=39\ \GeV$ and 62.4 GeV analyses used $3.5 \times 10^8$ and $7 \times 10^8$ events, respectively. 
The Beam-Beam counters were used to determine centrality. The number of participants (\Npart) and the number of binary collisions (\Ncoll) were calculated using the Glauber model. In both analyses neutral pions are observed through the $\piz\rightarrow\gamma\gamma$ decay channel. 
Photons are identified with the Electromagnetic Calorimeter (EMCal, $|\eta|<0.35$) \cite{Aphecetche:2003zr}. 
Only the Lead Scintillator part (PbSc, $3/4\pi$ in azimuth) was used in year 2010 analysis, while year 2007 high-statistics analysis used the Lead Glass (PbGl, $1/4\pi$) in addition. 
Extraction of the  \piz yield is detailed in~\cite{Adare:2012uk,Adare:2012wg}. 
The spectra are well described by a power-law function $f(\pT)=A\pT^{-n}$ at $\pT>4\ \GeV/c$, with the fitted slope parameters being $n_{39}=13.4\pm0.08$, $n_{62}=10.4\pm0.03$ and $n_{200}=8.06\pm0.01$ for $\snn=39$, 62.4 and 200 GeV minimum bias Au+Au collisions respectively. These values are comparable to the corresponding values from p+p, $n^{pp}_{39}=13.59\pm0.21$, $n^{pp}_{62}=9.82\pm0.18$ and $n^{pp}_{200}=8.22\pm0.09$.

Year 2007 $\snn=200\ \GeV$ $\Raa(\pT)$ was calculated point-by-point using p+p data from year 2005 run. At $\snn=62.4\ \GeV/c$, \piz cross section in p+p was measured in year 2006 run up to $\pT<7\ \GeV/c$. The reference was extrapolated into the range $7<\pT<10\ \GeV/c$, based on a power-law fit on the $4.5<\pT<7\ \GeV/c$ p+p data. For \snn=39 GeV, p+p data from Fermilab E706~\cite{Apanasevich:2000eq} were used, with correction for the different rapidity ranges using PYTHIA. Both methods import a 20\% uncertainty at high \pT.
Decay photons from high \pT pions have a small opening angle and, given the granularity of the EMCal, cannot be reconstructed separately. This "cluster merging" effect is negligible below $\pT<12\ \GeV$, but $\sim$50\% of $\piz$s are lost at \pt=16 GeV, yielding a 28\% error.
Other main sources of systematic uncertainties on the pion production are PID efficiency (3-5\%), yield extraction (3-5\%), conversion (4\%), acceptance (1-2\%) and off-vertex \piz{}s (1.5\%).
The reaction plane was determined by a combination of Muon Piston Calorimeter (MPC,  $3.1<|\eta|<3.9$) \cite{Adcox:2003zp} and the inner ring of the RxNP plastic scintillator ($1.5<|\eta|<2.8$) \cite{Richardson:2010hm}. The \piz{}s were classified into six bins depending on the emission angle with respect to the event plane, in order to measure $\Raa(\pT,\Dphi)$. An unfolding correction was applied due to finite event plane resolution.

\section{Evolution of \Raa and \Sloss with collision energy and centrality}
%
\begin{figure}[!t]
\centering
\vspace{-2mm}
\includegraphics[height=5.5cm,width=6cm]{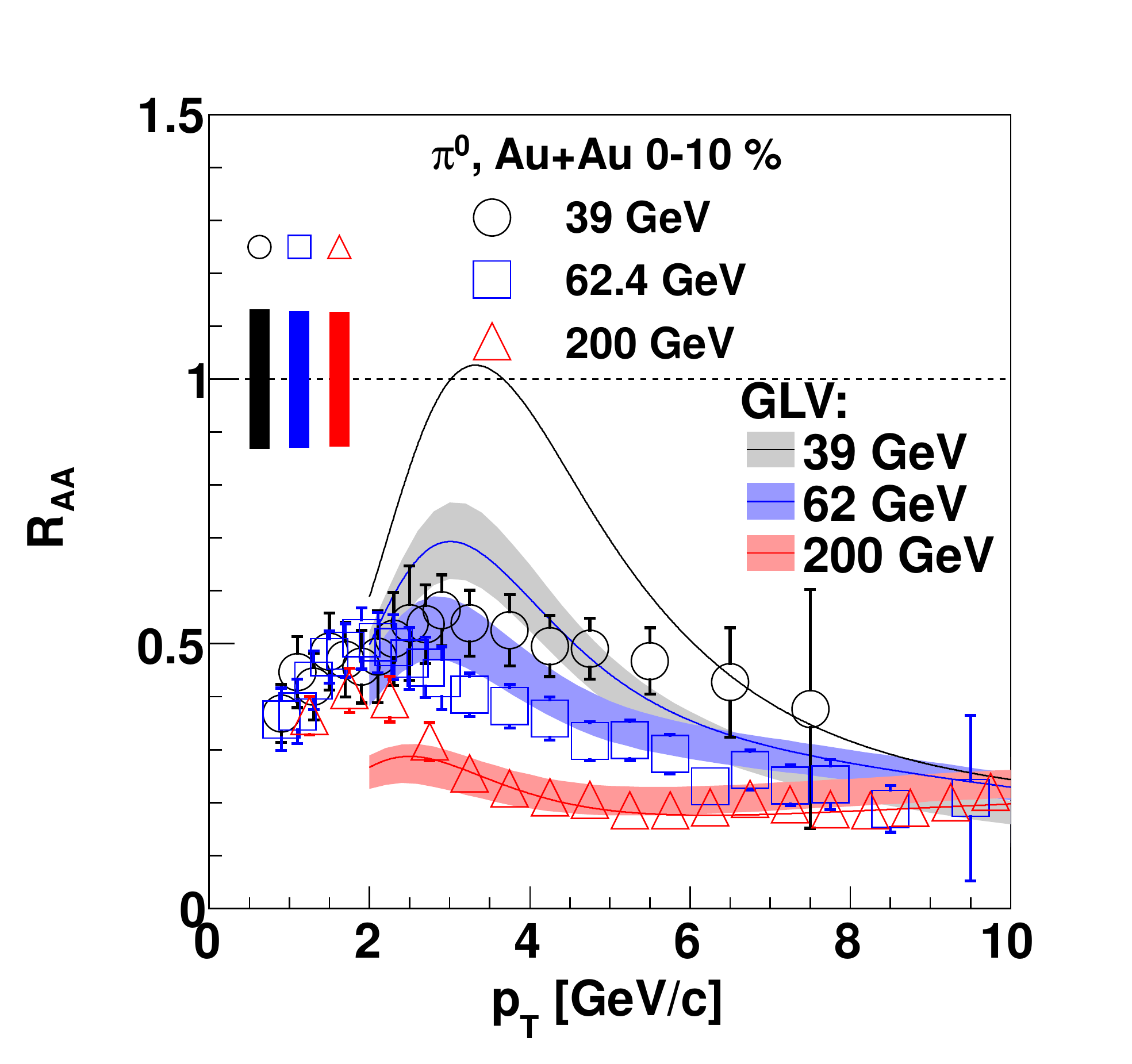}
\includegraphics[height=5.5cm,width=6cm]{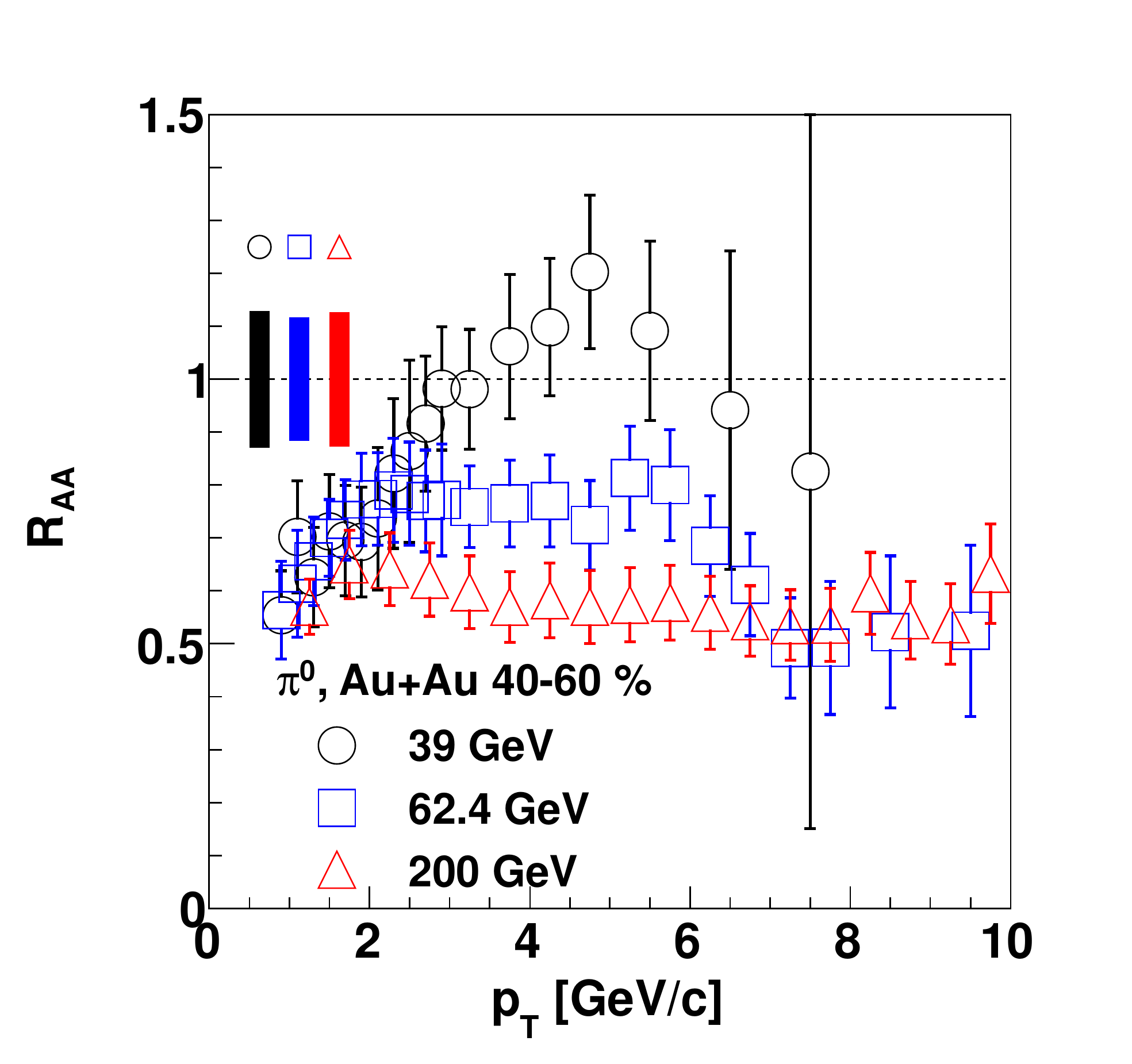}
\vspace{-2mm}
\caption{\Raa of \piz in
most central (0--10\%, left) and
mid-peripheral (40--60\%, right) Au+Au collisions. Error bars are the quadratic sum of statistical and \pT-correlated systematic uncertainties, including systematics from the p+p reference. Boxes around 1 are the quadratic sum of overall normalization and \Ncoll uncertainties. The continuous lines and the bands are pQCD calculations as described in the text.}
\label{fig:raa_pt_escan}
\end{figure}
\begin{figure}[!t]
\begin{minipage}[c]{0.52\linewidth}
\centering
\vspace{0mm}
\includegraphics[viewport=0 39 520 234,clip,width=7cm]{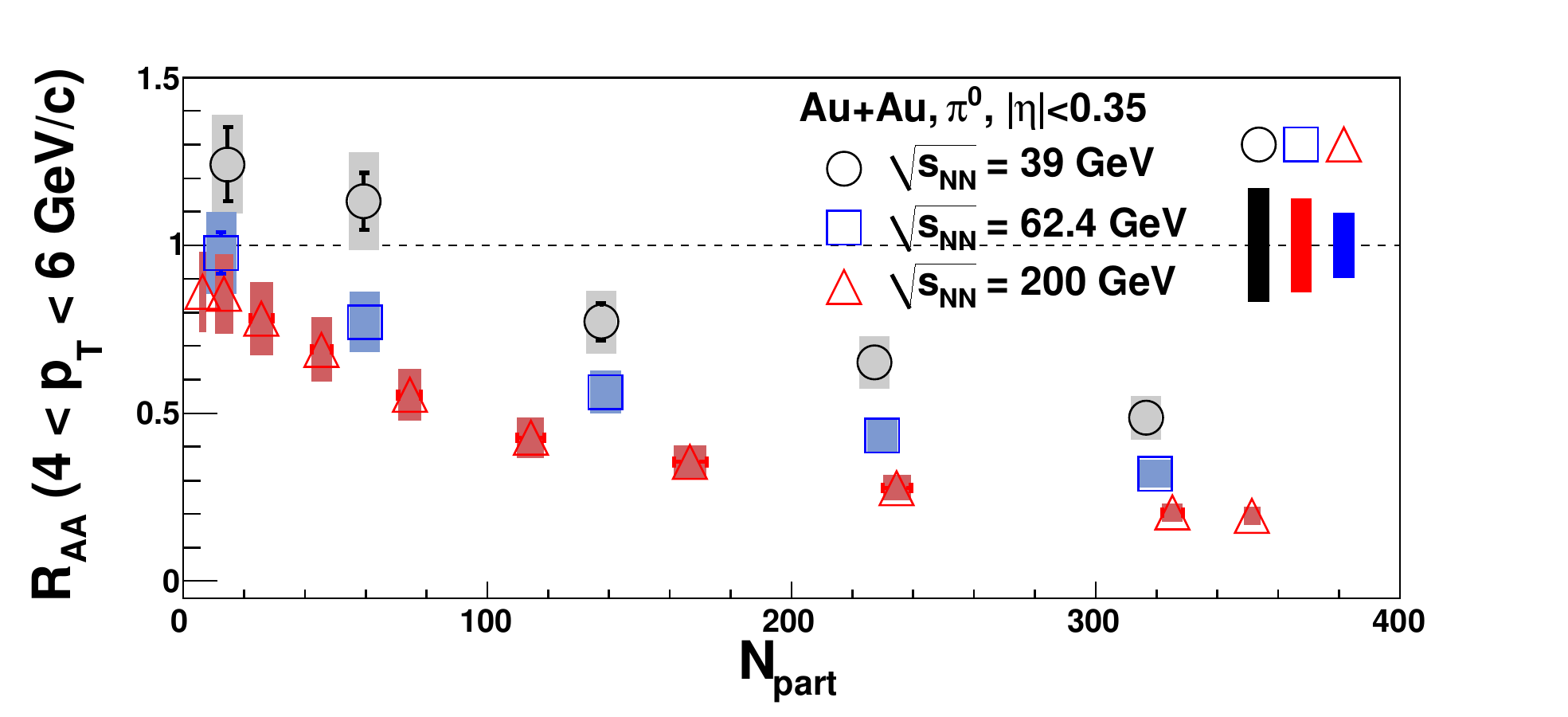}

\includegraphics[viewport=0 0 520 234,clip,width=7cm]{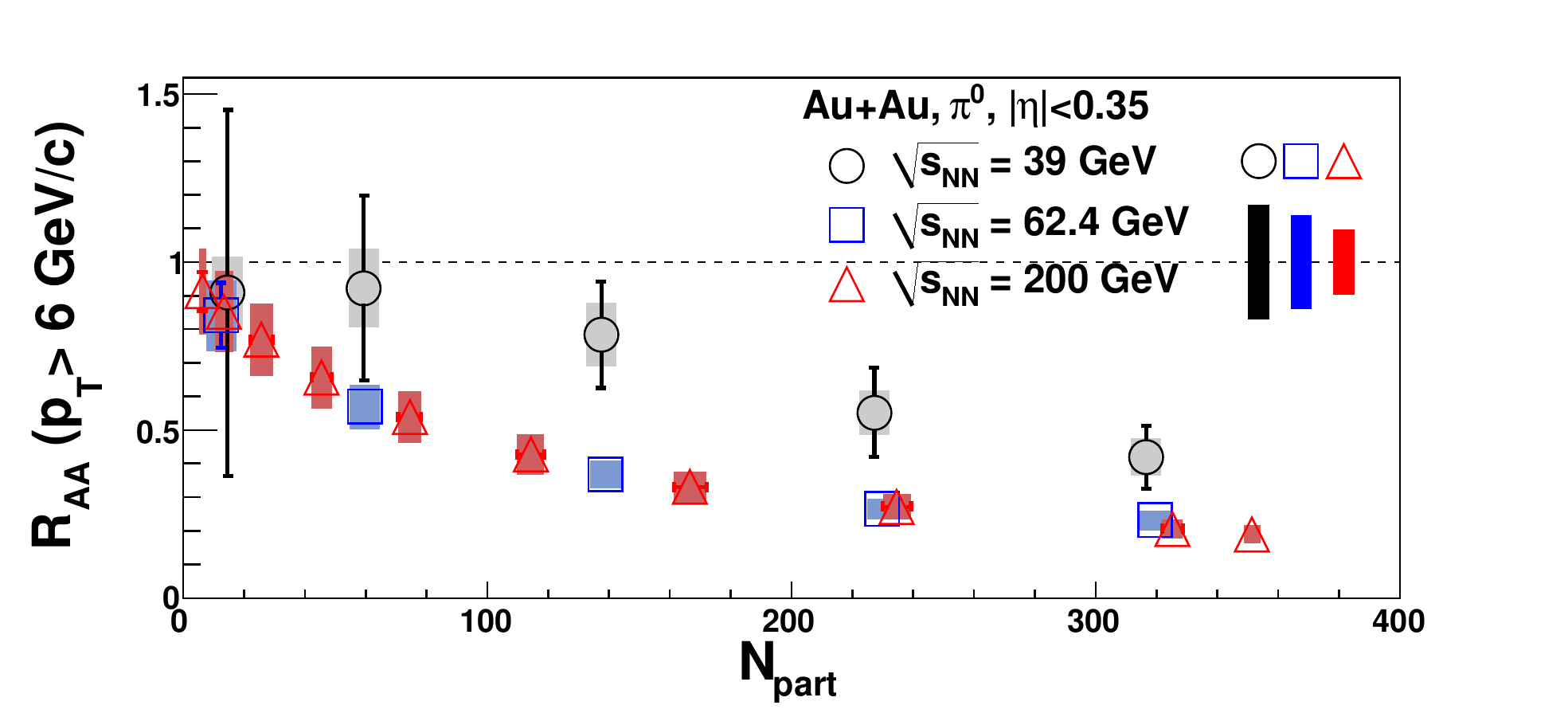}
\vspace{-2mm}
\caption{Nuclear modification factor averaged for $4<\pT<6\ \GeV$ (upper panel) and $\pT>6\ \GeV$ (lower panel). Error bars are statistical, boxes are the sum of \pT-uncorrelated and \Ncoll errors. Boxes around 1 are of \pT-correlated and overall normalization errors and errors from the p+p reference.}
\label{fig:raa_npart_escan}
\end{minipage}
\hspace{0.03\linewidth}
\begin{minipage}[c]{0.43\linewidth}
\centering
\vspace{-2mm}
\includegraphics[height=5.5cm,width=6cm]{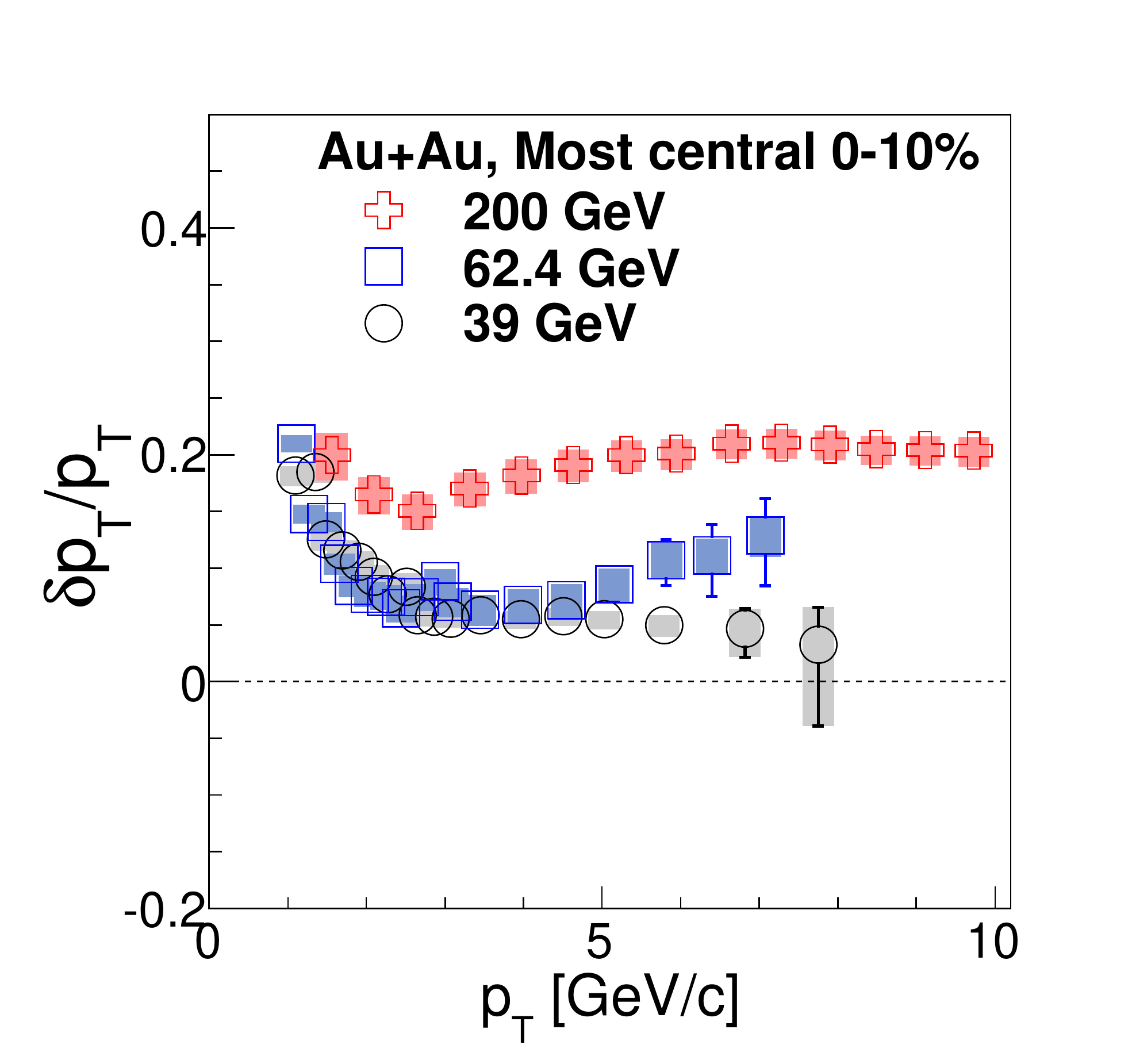}%
\caption{%
Fractional momentum shift \Sloss between Au+Au and scaled p+p collisions for central collisions at c.m.s energies of \snn=39, 62.4 and 200 GeV as a function of the Au+Au \pT.
}
\label{fig:sloss}
\end{minipage}
\end{figure}
The \piz nuclear modification factors $\Raa(\pT)$, measured by PHENIX in $\snn=39$, 62.4 and 200 GeV Au+Au reactions ~\cite{Adare:2008ad,Adare:2012uk} are shown in Fig.~\ref{fig:raa_pt_escan}. Data from the 10\% most central collisions (left panel) exhibit a strong suppression at all three energies. The mid-peripheral data (40\%-60\% centrality, right panel) at 62.4 and 200 GeV is suppressed in the whole \pT range, while at 39 GeV suppression is present at low \pT values, and rises slightly, though not significantly, above unity from $\pT>3\ \GeV$. 
In Fig~\ref{fig:raa_pt_escan}, pQCD-based calculations of~\cite{Sharma:2009hn} are shown as continuous lines. Although the model successfully describes the 200 GeV Au+Au and Cu+Cu data, it fails at 39 and 62.4 GeV. A calculation with reduced Cronin-effect and with the energy loss varied by $\pm10\%$, represented by shaded bands, is still unable to reproduce the trend in \pT observed in 39 and 62.4 GeV data.
The \piz \Raa as a function of $\Npart$ is shown in the ranges of $4<\pT<6\ \GeV/c$ and $\pT>6\ \GeV/c$ in Fig.~\ref{fig:raa_npart_escan}. Suppression in the higher \pT range is the same for $\snn=200\ \GeV$ and 62.4 GeV collisions, while the 39 GeV data departs from the trend. Note that there is a remarkable \Npart dependence at all energies including 39 GeV in both \pT ranges (also see Fig.~\ref{fig:raa_pt_escan}), and that \Raa{}s for all three energies converge at high \pT.
These observations, along with the radical changes of the slopes of the \pT{}-spectra with c.m.s. energies and the scaling violation of the \xt{}-scaling effective exponent~\cite{Adare:2012uk}, support the conclusion that jet quenching will be masked up to higher \pT, and hard scattering as a source of particles at a given \pT becomes completely dominant only at higher transverse momentum.

%
The p+p and Au+Au spectra have similar power law tails that can be fitted simultaneously with the function form $f(\pT)=A(\pT+\delta\pT)^{-n}$, where $\delta\pT$ is the horizontal shift between the two spectra. Fractional momentum shift $\Sloss=\delta\pT/\pT$ is assumed to correspond to the energy loss of partons within the medium. \Sloss is shown for central events at three different c.m.s. energies in Fig.~\ref{fig:sloss}. Although the naive picture that a large suppression in \Raa corresponds to a large \Sloss globally holds, it is to be noted that, contrary to the \Raa, \Sloss in 39 and 62.4 GeV is the same within errors below $\pT<6\ \GeV$, while in 200 GeV there is a higher momentum shift in the whole range. 

\section{Collision geometry dependence of \piz production}
\begin{figure}[!t]
\centering
\includegraphics[height=5cm]{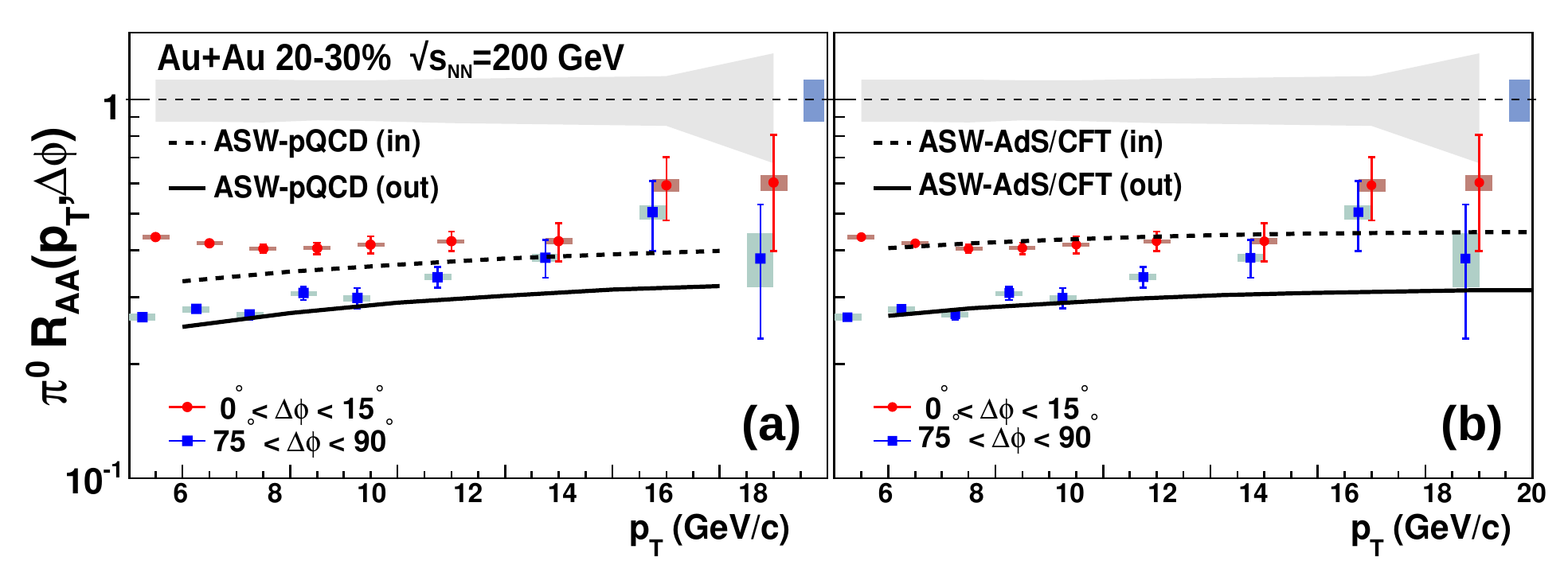}
\vspace{-2mm}
\caption{The data points are $\Raa(\Dphi)$ in 20--30\% centrality as a function of \pt for in-plane (dots, upper points) and out-of-plane $\piz$ (squares, lower points), compared to pQCD-based ASW~\cite{Salgado:2003gb} (a) and ASW using AdS/CFT correspondence~\cite{Marquet:2009eq} (b). The dotted and solid lines are the in-plane and out-of-plane predictions, respectively. The shaded bands around 1 are the systematic uncertainty of the normalizing $\phi$-integrated \Raa . Boxes on the right side of each panel around 1 show global uncertainties.}
\label{fig:raaphi}
\end{figure}
 
There are several theoretical models for the energy loss of the high-\pT particles within the medium, such as the Arnold-Yaffe formalism (AMY)~\cite{Arnold:2001ba}, the Higher Twist approach (HT)~\cite{Wang:2001ifa}, the pQCD-based Armesto-Salgado-Wiedemann approach (ASW)~\cite{Salgado:2003gb}, or the ASW using the AdS/CFT correspondance~\cite{Marquet:2009eq}. Since the azimuthal asymmetry of the overlap region develops into an anisotropy of the created medium, the average pathlength a high-\pT hadron traverses in the medium will be correlated to the azimuthal angle \Dphi between the path of the hadron and the reaction plane. 
The 2007 \snn=200 GeV Au+Au data on azimuthal anisotropy of \piz production extends the previous analysis of data from 2004 run~\cite{Afanasiev:2009aa} up to \pt=20 GeV/$c$, and provides a stronger discrimination power on the models. On Fig.~\ref{fig:raaphi} the nuclear modification factor for the in-plane ($0^\circ<\Dphi<15^\circ$) and out-of-plane ($75^\circ<\Dphi<90^\circ$) \piz{}s are plotted and compared to the ASW scenario based on pQCD as well as the AdS/CFT correspondance. While the out-of-plane $\Raa(\pT,\Dphi\approx\pi)$ is acceptably described by all models, measured in-plane $\Raa(\pT,\Dphi\approx 0)$ clearly favours the ASW scenario with AdS/CFT correspondence over the pQCD-based models. The pQCD-based AMY~\cite{Arnold:2001ba} and HT~\cite{Wang:2001ifa} are also not supported by the data~\cite{Adare:2012wg}.

\section{Summary}

New results from the 2010 RHIC low energy program show a substantial suppression of neutral pions in central Au+Au collisions at both \snn=39 and 62.4 GeV c.m.s. energies. At $\pT>6\ \GeV/c$ the 62.4 GeV and 200 GeV data follow the same suppression pattern. On the other hand, otherwise succesful pQCD predictions do not describe the 39~GeV data. The fractional momentum shifts \Sloss in 62.4 or 39 GeV collisions are similar below $\pT<6\ \GeV/c$, while \Sloss is substantially higher for 200 GeV than 62.4 or 39 GeV collisions, implying a higher fractional energy loss. These observations indicate that initial state effects may play a dominant role at smaller c.m.s. energies and at lower \pT, that are able to mask the effect of suppression, while the hard processes become the dominant sources of hadrons above $\pT\approx 6\ \GeV/c$. 

The  azimuthal dependence of the nuclear modification factor \Raa is strongly correlated with the (approximately elliptical) geometry of the overlap region. The dependence of \Raa on the reaction plane, determined up to \pt{}=20 GeV/$c$ from 2007 high-luminosity \snn=200 GeV Au+Au data provides great selectivity among theories, and favours the ASW scenario with AdS/CFT correspondence over the pQCD-based models.



\begin{thebibliography}{00}

\bibitem{Adcox:2001jp} 
  K.~Adcox {\it et al.} (PHENIX collaboration), 
  Phys.\ Rev.\ Lett.\  {\bf 88}, 022301 (2002).

\bibitem{Adler:2003qi} 
  S.~S.~Adler {\it et al.} (PHENIX collaboration), 
  Phys.\ Rev.\ Lett.\ {\bf 91}, 072301 (2003).

\bibitem{Adler:2003ii} 
  S.~S.~Adler {\it et al.} (PHENIX collaboration),
  Phys.\ Rev.\ Lett.\ {\bf 91}, 072303 (2003).

\bibitem{Adare:2008ad} 
  A.~Adare {\it et al.} (PHENIX collaboration),
  Phys.\ Rev.\ Lett.\ {\bf 101}, 162301 (2008).

\bibitem{Adare:2012uk} 
  A.~Adare {\it et al.} (PHENIX collaboration),
  arXiv:1204.1526 [nucl-ex], submitted to Phys.\ Rev.\ Lett.\ (2012).
  
\bibitem{Adare:2012wg} 
  A.~Adare {\it et al.} (PHENIX collaboration), 
  arXiv:1208.2254 [nucl-ex], submitted to Phys.\ Rev.\ C (2012).

\bibitem{Adcox:2003en} 
  K.~Adcox {\it et al.} (PHENIX collaboration),
  Nucl.\ Instrum.\ Meth.\ A {\bf 497}, 263 (2003).

\bibitem{Aphecetche:2003zr} 
  L.~Aphecetche {\it et al.}  (PHENIX collaboration),
  Nucl.\ Instrum.\ Meth.\ A {\bf 499}, 521 (2003).

\bibitem{Apanasevich:2000eq} 
  L.~Apanasevich {\it et al.},
  Phys.\ Rev.\ D {\bf 63}, 014009 (2001).

\bibitem{Adcox:2003zp} 
  K.~Adcox {\it et al.} (PHENIX collaboration),
  Nucl.\ Instrum.\ Meth.\ A {\bf 499}, 489 (2003).

\bibitem{Richardson:2010hm} 
  E.~Richardson {\it et al.} (PHENIX collaboration),
  Nucl.\ Instrum.\ Meth.\ A {\bf 636}, 99 (2011).

\bibitem{Sharma:2009hn} 
  R.~Sharma, I.~Vitev and B.~-W.~Zhang,
  Phys.\ Rev.\ C {\bf 80}, 054902 (2009). 
 
%
\bibitem{Afanasiev:2009aa} 
  S.~Afanasiev (PHENIX collaboration) {\it et al.},
  Phys.\ Rev.\ C {\bf 80}, 054907 (2009).

\bibitem{Arnold:2001ba} 
  P.~B.~Arnold, G.~D.~Moore and L.~G.~Yaffe,
  JHEP {\bf 0111}, 057 (2001).

\bibitem{Wang:2001ifa} 
  X.~-N.~Wang and X.~-f.~Guo,
  Nucl.\ Phys.\ A {\bf 696}, 788 (2001).

\bibitem{Salgado:2003gb} 
  C.~A.~Salgado and U.~A.~Wiedemann,
  Phys.\ Rev.\ D {\bf 68}, 014008 (2003).
  
\bibitem{Marquet:2009eq} 
  C.~Marquet and T.~Renk,
  Phys.\ Lett.\ B {\bf 685}, 270 (2010)

\end{thebibliography}
\end{document}